\documentclass[prd,aps,superscriptaddress,preprintnumbers,floatfix,showpacs,byrevtex,twocolumn]{revtex4}
\usepackage{graphicx}
\usepackage{amsmath}
\newcommand{\be}{\begin{equation}}
\newcommand{\ee}{\end{equation}}

\newcommand{\slh}{\!\!\!\slash}

\newcommand{\tr}{\text{Tr}}

\newcommand{\eref}[1]{Eq.~(\ref{#1})}

\newcommand{\dcsb}{D$\chi$SB}

\begin{document}

\preprint{
\vbox{
\hbox{ADP-10-18/T714}
}}

\title{Role of center vortices in chiral symmetry breaking in SU(3)
gauge theory}

\author{Patrick~O.~Bowman}
\affiliation{Centre for Theoretical Chemistry and Physics,
  and Institute of Natural Sciences, Massey University (Albany),
  Private Bag 102904, North Shorce MSC, New Zealand}

\author{Kurt~Langfeld}
\affiliation{School of Maths \& Stats, University of Plymouth,
  Plymouth, PL4 8AA, England}

\author{Derek~B.~Leinweber}
\affiliation{Centre for the Subatomic Structure of Matter
  (CSSM), School of Chemistry \& Physics, University of Adelaide 5005,
  Australia}

\author{Andr\'e~Sternbeck}
\affiliation{Centre for the Subatomic Structure of Matter
  (CSSM), School of Chemistry \& Physics, University of Adelaide 5005,
  Australia}
\affiliation{Institut f\"ur Theoretische Physik, Universit\"at Regensburg,
  D-93040 Regensburg, Germany}

\author{Lorenz von Smekal}
\affiliation{Centre for the Subatomic Structure of Matter
  (CSSM), School of Chemistry \& Physics, University of Adelaide 5005,
  Australia}
\affiliation{Institut f\"ur Kernphysik, Technische Universit\"at
  Darmstadt, D-64289 Darmstadt, Germany}

\author{Anthony G. Williams}
\affiliation{Centre for the Subatomic Structure of Matter
  (CSSM), School of Chemistry \& Physics, University of Adelaide 5005,
  Australia}

\begin{abstract}
  We study the behavior of the AsqTad quark propagator in Landau gauge
  on SU(3) Yang-Mills gauge configurations under the removal of center
  vortices.  In SU(2) gauge theory, center vortices have been observed
  to generate chiral symmetry breaking and dominate the infrared
  behavior of the quark propagator.  In contrast, we report a weak
  dependence on the vortex content of the gauge configurations, including
  the survival of dynamical mass generation on configurations
  with vanishing string tension.
\end{abstract}


\maketitle

\section{INTRODUCTION}

The strong nuclear force has two key features: the dynamical breaking of chiral
symmetry (\dcsb) and the confinement of color-charged states.  It is tempting to
attribute these two phenomena to a single underlying mechanism, an idea
supported by finite-temperature studies where the deconfinement and chiral
restoration transitions are observed to occur at coincident
temperatures~\cite{Laermann:2003cv}.
Low-lying modes of the quark operator, known to dominate \dcsb, are also
correlated with the finite-temperature transition of the Polyakov loop, and
hence confinement~\cite{Gattringer:2006ci,Bruckmann:2006kx,Synatschke:2007bz,Synatschke:2008yt}.

Over the recent past, evidence has been accumulated by means of lattice gauge
theories that both phenomenon are caused by certain low energy degrees
of freedoms. In specific gauges, these degrees of freedom appear as
colour-magnetic monopoles~\cite{Kronfeld:1987ri,Suzuki:1989gp,Suganuma:1993ps}
or center vortices~\cite{DelDebbio:1996mh,DelDebbio:1998uu,Greensite:2003bk}.
The idea that center fluxes disorder Wilson loops, and therefore lead to
confinement, is an old one~\cite{'tHooft:1977hy,'tHooft:1979uj}
and over the last couple of decades a
great deal of work has been done in Lattice Gauge Theory on such objects,
principally in SU(2) Yang-Mills theory. It turned out to be difficult
to define the vortex content of Yang-Mills theory in a physically sensible
way. It took until the late nineties until a successful definition was
given~\cite{DelDebbio:1996xm} and the relevance of vortices
in the continuum limit was established~\cite{Langfeld:1997jx}.
The recovery of the string tension from ``vortex-only'' SU(2) gauge
configurations (i.e., $\text{Z}_2$ projected from SU(2))
was shown~\cite{DelDebbio:1996mh,DelDebbio:1998uu,Greensite:2003bk},
the finite temperature deconfinement transition was understood
in terms of vortex
properties~\cite{Langfeld:1998cz,Engelhardt:1999fd,Langfeld:2003zi}
and a connection to \dcsb\ was
discovered~\cite{deForcrand:1999ms,Langfeld:2003ev,Gattnar:2004gx}.

The use of Landau gauge Green's functions as probes of \dcsb\ and
confinement is an active area of research (see, e.g.,
\cite{Alkofer:2000wg, Fischer:2006ub} for a review).  It is known, for
example, gluon propagator violates spectral positivity, which is
consistent with gluon
confinement~\cite{Cucchieri:2004mf,Sternbeck:2006cg,Bowman:2007du}.
In the quark propagator the Dirac scalar part, related at large
momenta to the perturbative running mass, is enhanced at low momenta,
even in the chiral limit~\cite{Bowman:2005zi,Roberts:2007ji}: a
demonstration of \dcsb.  One feature of this approach is that it
allows one to make statements about light quarks, as opposed to the
static potential of the Wilson loop.  In SU(2) gauge theory the
infrared properties of the quark propagator were found to be dominated
by center vortices~\cite{Bowman:2008qd,Hollwieser:2008tq}.
Unfortunately, the vortex picture for the gauge group SU(3) is less
clear: while vortex removal eliminates the linear rise of the static
quark potential at large distances, the string tension of vortex only
configurations falls short by roughly a factor
$2/3$~\cite{Langfeld:2003ev,Cais:2007bm,Cais:2008za}.

To gain further insights into the SU(3) vortex picture, we here
investigate the SU(3) quark propagator under the removal of center vortices.
We will find that mass generation remains intact even after removing
center vortices, while the string tension vanishes as expected.

\section{Center Vortices}

We will identify center vortices in SU(3) Yang-Mills lattice gauge
configurations using standard methods.  Having generated gauge
configurations, we will rotate them to Direct Maximal Center gauge
then project the gauge links onto the nearest center element.  Each
configuration can then be decomposed into two pieces: the center
element and ``the rest''.  An appealing result of such a decomposition
would be the identification of separate short and long-ranged pieces,
such as seen in SU(2) gauge theory~\cite{Bowman:2008qd}; that is, that
this decomposition corresponds to a separation of infrared
(vortex-only) and ultraviolet (vortex-removed) physics.  Finally, to
study the propagators, the vortex-only and vortex-removed
configurations are rotated to Landau gauge and the quark propagators
calculated.

A statistical ensemble of lattice gauge configurations is generated using the
L{\"u}scher-Weisz~\cite{Luscher:1984xn} mean-field improved action,
\begin{multline}
S_{\rm G} = \frac{5\beta}{3}
      \sum_{\rm{sq}}\frac{1}{3}{\cal R}e\ {\rm{tr}}(1-U_{\rm{sq}}(x))
\nonumber \\
 - \frac{\beta}{12u_{0}^2}
      \sum_{\rm{rect}}\frac{1}{3}{\cal R}e\ {\rm{tr}}(1-U_{\rm{rect}}(x))\ ,
\label{gaugeaction}
\end{multline}
where $U_{\rm{sq}}(x)$ is the plaquette and $U_{\rm{rect}}(x)$
denotes the rectangular $1\times2$ and $2\times1$ loops.  For the
tadpole improvement factor we employ the gauge-invariant plaquette measure
\be
u_0 = \left( \frac{1}{3}{\cal R}e \, {\rm{tr}}\langle U_{\rm{sq}}\rangle
      \right)^{1/4}\ .
\label{uzero}
\ee

\subsection{Maximal Center Gauge}
\label{sec:vortices}

In order to identify the center fluxes of a given lattice configuration it is
common to use gauge fixing and center projection.
The center fluxes through an elementary plaquette are represented
by center link elements $Z_\mu (x)$ which take values in the center
group $Z_3 \subset SU(3)$:
$$
Z_\mu (x) \; = \; \exp \Bigl\{ i \frac{2\pi }{3} m_\mu (x) \Bigr\} \; ,
\; \; \; \;  m_\mu (x) \in \{-1,0,1 \} \; .
$$
It is  a non-trivial task to find a definition of the center links that is
sensible in the continuum limit.  The following definition has turned out
to be fruitful~\cite{DelDebbio:1996mh,Langfeld:1997jx,Langfeld:2003ev}:
\begin{equation}
\label{eq:min}
\sum _{x, \mu } \Bigl\lVert U_\mu^\Omega (x) \; - \;
Z_\mu (x) \Bigr\rVert \; \stackrel{ \Omega, Z_\mu (x) }{ \longrightarrow }
\; \text{min}.
\end{equation}
This has an intuitive interpretation:
After a suitable gauge transformation $\Omega (x)$, we look for
those center links $Z_\mu(x)$ that represent best a given link
$U_\mu (x)$. \eref{eq:min} implies that the overlap
between the gauged links and the center links is maximized:
\be
\sum _{x, \mu } {\cal R}e\ \Bigl[ \tr  U_\mu^\Omega (x)\; Z_\mu^\dagger (x)
\Bigr] \; \stackrel{ \Omega, Z_\mu (x) }{ \longrightarrow }
\; \mathrm{max}.
\label{eq:overl}
\ee
Hence, we will exploit the gauge degrees of freedom to bring
$U_\mu^\Omega (x)$ as close as possible to a center element.
Assuming that the deviations of $U_\mu^\Omega (x)$ from a center
element are small, one might approximately solve (\ref{eq:overl})
by setting
\begin{equation}
Z_\mu (x) \approx \frac{1}{3} \tr U_\mu ^\Omega (x) \; , \; \; \mathrm{or}
\; \;
Z_\mu (x) \approx \Bigl[ \frac{1}{3} \tr U_\mu ^{\dagger \, \Omega } (x)
\Bigr]^2 \; .
\end{equation}
One gauge condition for determining the gauge transformation $\Omega $ is,
\begin{eqnarray}
R_{\rm mes} = \sum _{x, \mu } \Bigl\vert \tr  U_\mu^\Omega (x) \Bigr\vert ^2
\; & \stackrel{ \Omega }{ \longrightarrow } &
\; \mathrm{max},
\label{eq:mesonic}
\end{eqnarray}
This gauge conditions specifies a particular Maximal Center Gauge, known in the
literature as the `mesonic' center
gauge~\cite{Faber:1999sq,Montero:1999by,Langfeld:2003ev}.

\subsection{Center Projection and Vortex Removal }

Once the optimal choice for the gauge transformation $\Omega (x)$
is obtained, the center links $Z_\mu (x)$ are obtained from the gauged links
$U^\Omega _\mu (x)$ by center projection. Decomposing a particular link,
\begin{equation}
\frac{1}{3} \tr U^\Omega _\mu (x) \; = \; r_\mu (x) \;
\exp \Bigl(i \varphi _\mu (x) \Bigr) \; ,
\end{equation}
where $r_\mu (x)$ is real and
$ \varphi _\mu (x) \in [-\pi, \pi )$, \eref{eq:overl} implies
that we locally maximize
\begin{equation}
\cos \left[ \varphi _\mu (x) \; - \; \frac{2 \pi }{3} m_\mu (x)
\right] \; \stackrel{m_\mu }{\longrightarrow } \;
\mathrm{max} \; .
\end{equation}
Hence, the integer $m_\mu (x) \in \{-1,0,1\}$ closest to
$3 \; \varphi _\mu (x)/2\pi $ is chosen. Once the center links
$Z_\mu (X)$ are obtained in this way, center fluxes $\phi _{\mu \nu }(x) $
are detected from the center plaquettes
\begin{eqnarray}
P_{\mu \nu }(x) &=& Z_\mu (x) \, Z_\nu (x+\mu) \,
Z^\dagger _\mu (x+\nu) \, Z^\dagger _\nu (x)
\nonumber \\
&=&   \exp \Bigl\{ i \frac{2\pi }{3} \phi_{\mu \nu } (x) \Bigr\} \; ,
\end{eqnarray}
where
$$
\phi_{\mu \nu } (x) \: \in \; \{-1,0,1\} \; .
$$
We say that a particular plaquette $(\mu, \nu ;x)$ is intersected
by nontrivial center flux if $\phi_{\mu \nu } (x) \not =0$.
It can be shown, using the $Z_3$ Bianchi identity, that
the set of plaquettes that carry non-trivial center flux form
closed surfaces on the dual lattice. These surfaces define the world
sheets of $Z_3$ vortices.
The theory without center fluxes (vortex-removed configurations) is defined
from the link elements
\be
\widetilde{U}_\mu (x) \; \equiv \; U^\Omega _\mu (x) \; Z^\dagger _\mu (x) \; .
\label{eq:urem}
\ee

\subsection{Numerical results }

\begin{figure*}[t]
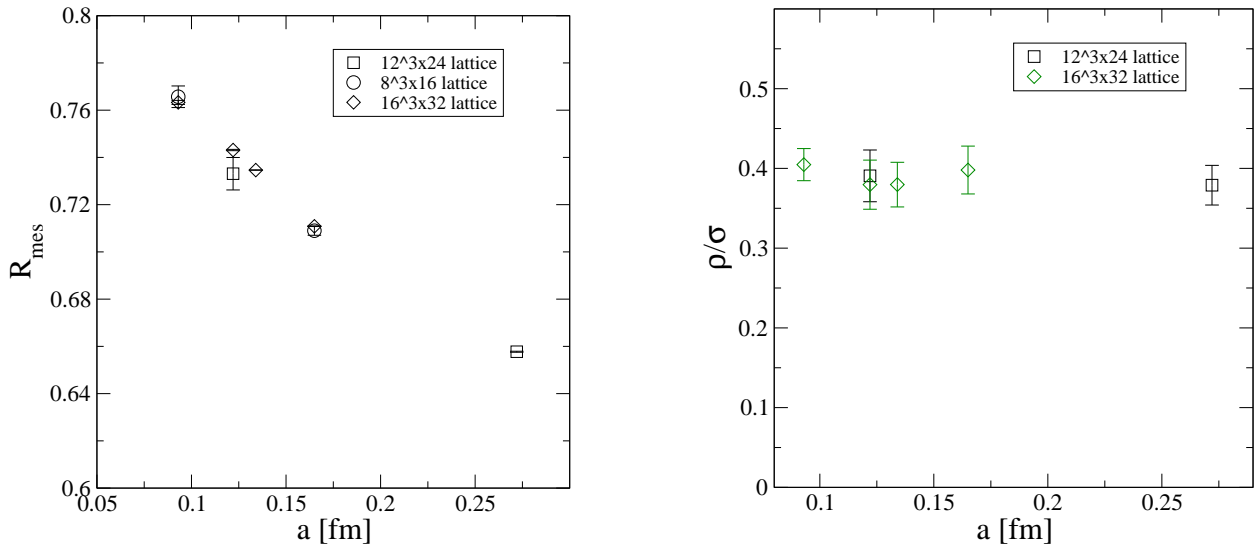

  \begin{center}
    \includegraphics[height=0.4\hsize]{./Rvsa.eps} \hspace{1.5cm}
    \includegraphics[height=0.4\hsize]{./rhoKvsa.eps}
    \caption{Left panel: The gauge fixing functional $R_{\rm mes}$
      after MCG gauge 
      fixing.  Right panel: the vortex area density $\rho $ in units of the
      string tension $\sigma $ as function of the lattice spacing $a$.
    }
    \label{fig:R_rho}
  \end{center}
\end{figure*}

The configurations are fixed to Maximal Center Gauge by maximizing the
gauge fixing functional (\ref{eq:mesonic}) with the help of a local
update algorithm. The algorithm is presented in detail
in~\cite{Montero:1999by}. Lattice sizes and simulation parameters
are listed in Table~\ref{tab:lat}.

In Fig.~\ref{fig:R_rho}, we show the final value of the gauge fixing
functional $R_\mathrm{mes}$ for several values of the lattice spacing $a$.
Generically, increasing values for $R_\mathrm{mes}$ are obtained for decreasing
lattice spacing.  This indicates that the overlap of the full configurations
with pure center ones increases towards the continuum limit.

\begin{table}[hbt]
\begin{tabular}{c|c|c|c|c|c|c}
$\beta$ & Volume & $N_{\mathrm{con}}$ & $a\sqrt{\sigma}$ & $a_{\sigma}$
[fm] & $\rho a^2 $ & $\rho /\sigma $ \\\hline
4.10    & $12^3\times 24$ & 15            & 0.611(20) & 0.272(9) & 0.1414(4) &
0.379(25) \\
4.38    & $16^3 \times 32$ & 100          & 0.368(5)  & 0.165(3) & 0.0539(2)
& 0.398(30)    \\
4.53    & $16^3\times 32$ & 100           & 0.299(11) & 0.134(5) & 0.0339(2) &
0.380(28) \\
4.60    & $16^3\times 32$ & 100           & 0.272(11) & 0.122(5) & 0.0281(2) &
0.380(31) \\
4.60    & $12^3\times 24$ & 15            & 0.272(11) & 0.122(5) & 0.0289(5) &
0.391(32) \\
4.80    & $16^3\times 32$ & 100           & 0.207(5)  & 0.093(2) & 0.0173(2) & 0.404(20)
\end{tabular}
\caption[Simulation parameters]{Simulation parameters $\beta$, volumes, string
  tension $a\sqrt{\sigma}$, lattice spacings $a$ and vortex densities. The
  values for the lattice spacings for the $16^3\times 32$
  lattices have been obtained by using 50 configurations each. For the
  small $\beta=4.60$ lattice estimates are taken from the
  larger lattice.}\label{tab:lat}
\end{table}

If $\rho $ denotes the planar vortex area density then the quantity
$\rho a^2$ can be interpreted as the probability that
a given plaquette carries non-trivial center flux.  We have calculated
$\rho a^2$ for several lattice spacings (see Table~\ref{tab:lat})
by counting the number of plaquettes with non-trivial center fluxes and
then dividing this number by the total number of plaquettes on the
lattice.
The interesting observation is that the planar vortex density, $\rho $,
is independent of the lattice spacing $a$ (Fig.~\ref{fig:R_rho}) and therefore
has a sensible continuum interpretation.
This behavior is in accordance with the behavior of the SU(2) vortex
density, and confirms earlier findings for the gauge group
SU(3)~\cite{Langfeld:2003ev}.

\begin{figure}[t]
  \begin{center}
    \includegraphics[height=0.8\hsize]{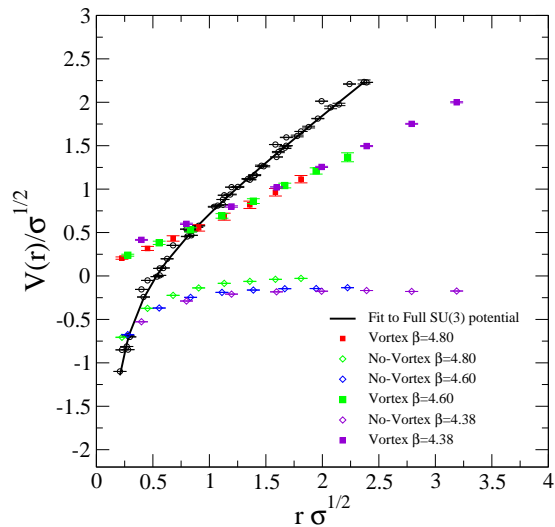}
    \caption{The quark anti-quark potential $V(r)$ as function of the
      quark anti-quark distance $r$ for full configurations (open circles),
      vortex-only configurations (full symbols) and vortex-removed
      configurations (open symbols).
    }
    \label{fig:potvor}
  \end{center}
\end{figure}

Now, we calculate the static quark anti-quark potential for ensembles with
full SU(3) links, for vortex-only configurations and for vortex-removed
configurations.
We observe that the vortex-removed configurations show no sign of a confining
potential. On the other hand, the vortex-only ensembles
give rise to around $60\%$ of the string tension. This confirms earlier
findings~\cite{Langfeld:2003ev,Cais:2007bm} and is in sharp contrast
to the case of SU(2) gauge theory: there, the vortices reproduce
a great deal of the full string tension~\cite{Greensite:2003bk}.

\section{Quark Propagator on the Lattice}
\label{sec:quarkprop}

In a covariant gauge in the continuum, Lorentz invariance allows us to
decompose the full quark propagator into Dirac vector and scalar
pieces.  In momentum space, the renormalized Euclidean space quark
propagator has the form
\begin{eqnarray}
S(\zeta;p)=\frac{1}{i {p \slh} A(\zeta;p^2)+B(\zeta;p^2)}
=\frac{Z(\zeta;p^2)}{i{p\slh}+M(p^2)}\, ,
\label{ren_prop}
\end{eqnarray}
where $\zeta$ is the renormalization point.

When the quark-gluon interactions are turned off, the quark propagator
takes its tree-level form
\begin{equation}
S^{(0)}(p)=\frac{1}{i{p\slh}+m} \, ,
\end{equation}
where $m$ is the bare quark mass.  When the interactions with the
gluon field are turned on we have
\begin{equation}
S^{(0)}(p) \to S^{\rm bare}(a;p) = Z_2(\zeta;a) S(\zeta;p) \, ,
\label{tree_bare_ren}
\end{equation}
where $a$ is the regularization parameter (i.e., the lattice spacing) and
$Z_2(\zeta;a)$ is the renormalization constant.  In the MOM scheme it
is chosen so as to ensure tree-level behavior at the renormalization point,
$Z(\zeta;\zeta^2)=1$.
Note that $M(p^2)$ is renormalization point independent,
i.e., since $S(\zeta;p)$ is multiplicatively renormalizable all of the
renormalization-point dependence is carried by $Z(\zeta;p^2)$.
For simplicity
of notation we suppress the $a$-dependence of the bare quantities.

In this work we use the AsqTad quark action~\cite{Orginos:1999cr} because of
its excellent scaling and rotational symmetry properties~\cite{Bowman:2005zi}.
The Dirac scalar and vector functions, $M(p^2)$
and $Z(p^2)$ are extracted from the propagator using the techniques described
in detail in Ref.~\cite{Bowman:2005vx}.

When analysing our results we will sometimes find it convenient to use a
``cylinder cut'' \cite{Leinweber:1998im}, where we select only data with
four-momentum lying near the four-dimensional diagonal. This is motivated by the
observation that for a given momentum squared, ($p^2$), choosing the
smallest momentum values of each of the Cartesian components, $p_\mu$,
should minimize finite lattice spacing artifacts.  By eliminating points most
likely to be affected by hyper-cubic lattice artifacts it is easier to draw
robust conclusions.

\section{The Influence of Center Vortices}

\begin{figure*}[t]
  \begin{center}
    \includegraphics[height=0.45\hsize,angle=90]{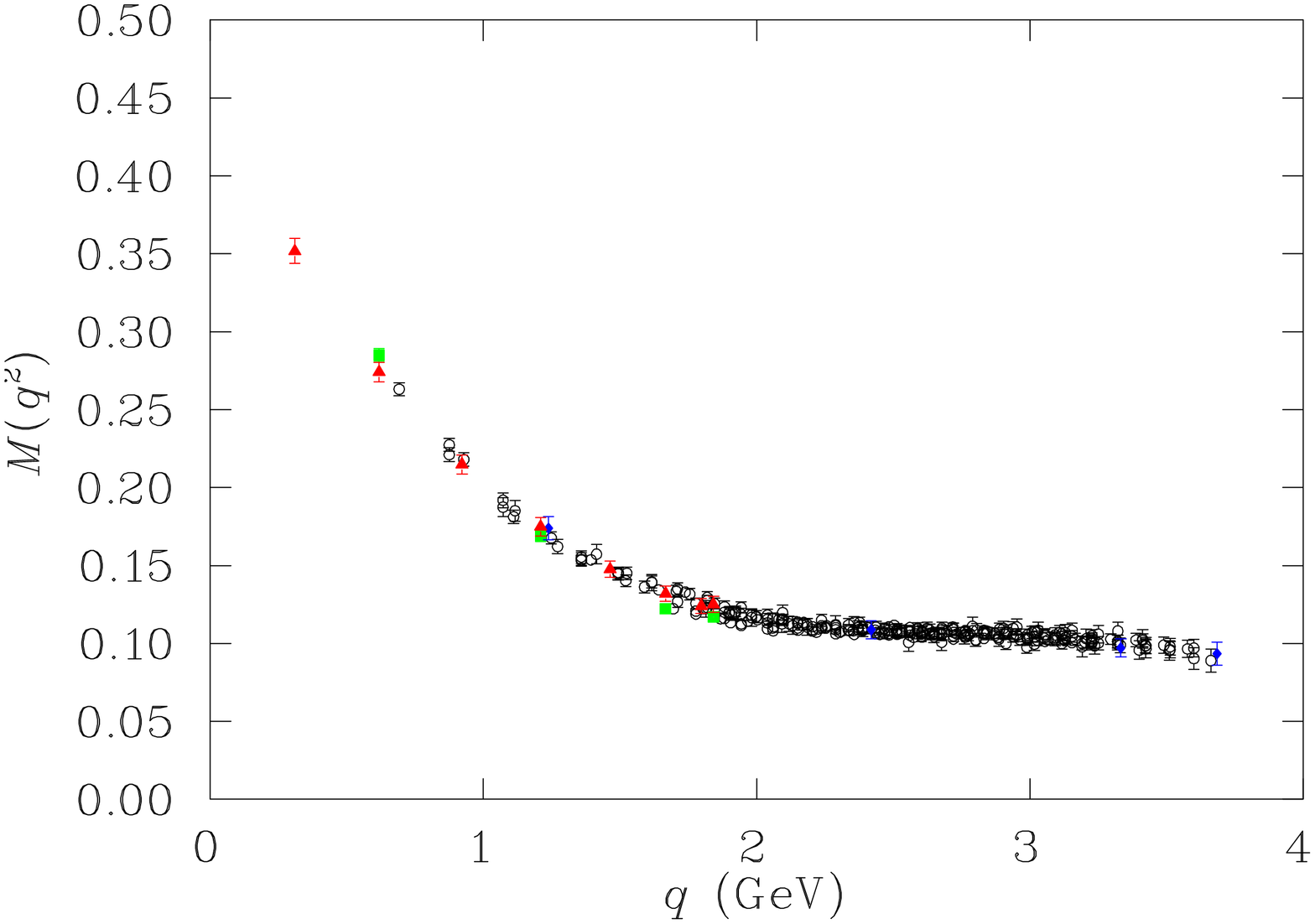}\hspace{5mm}
    \includegraphics[height=0.45\hsize,angle=90]{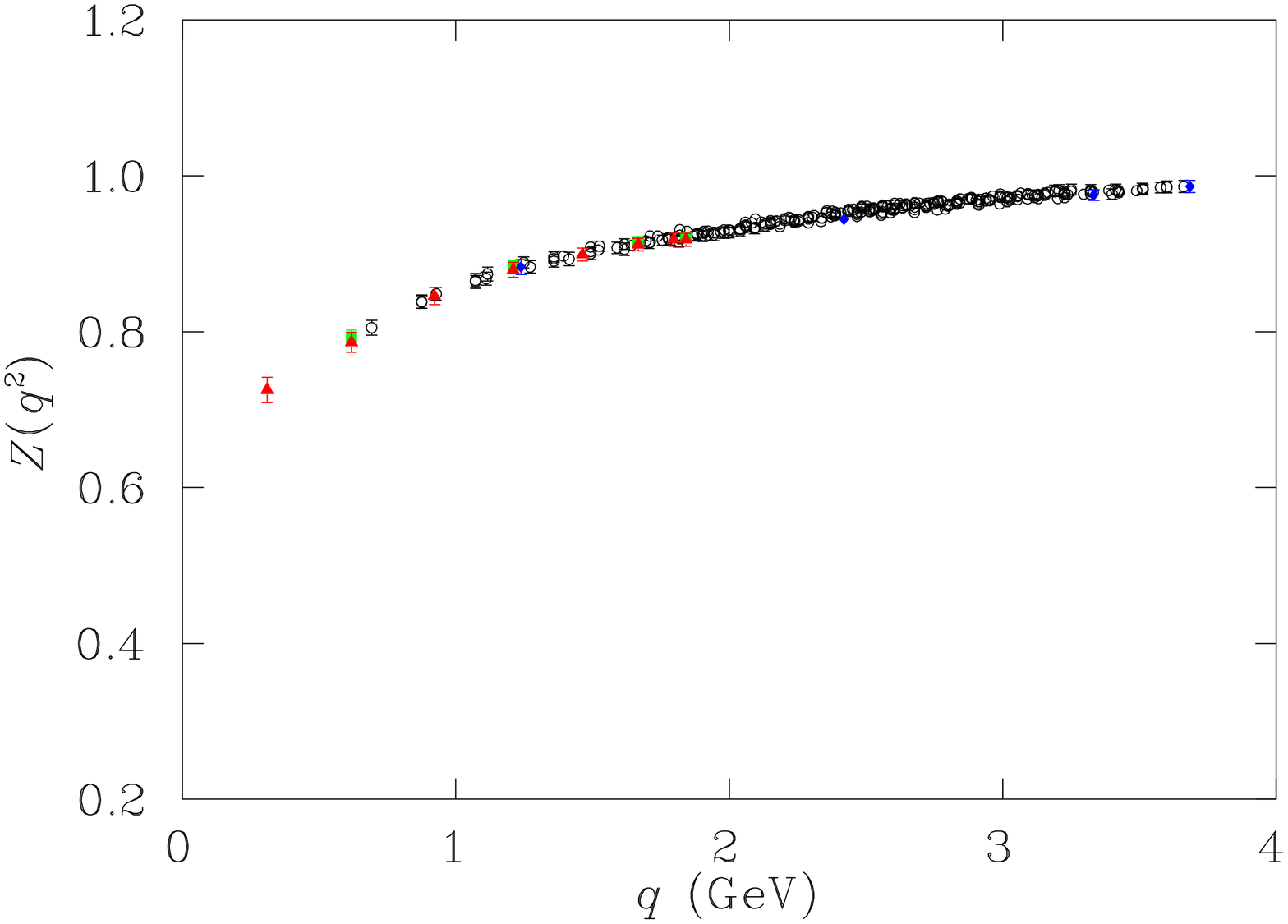}
    \caption{The Landau gauge quark propagator.  The left panel shows the mass
      function $M(q^2)$ and the right panel the wave-function renormalization
      function $Z(q^2)$ for $m_0 a = 0.048$.  The infrared enhancement of the
      mass function demonstrates \dcsb.}
    \label{M048}
  \end{center}
\end{figure*}

The Landau gauge quark propagator is calculated on the $16^3\times 32$
configurations at $\beta=4.60$.
The quark mass and wave-function renormalization functions of the original
untouched gauge configurations are illustrated in
Figure~\ref{M048}.  Here symbols are used to identify
momenta having a particular orientation within the lattice.  Triangles
denote momenta lying along the Cartesian time direction (the long dimension),
squares denote momenta oriented along one spatial Cartesian
direction, and diamonds denote momenta oriented along the lattice four
diagonal.  A comparison of triangles and squares is
useful in revealing finite volume effects at small momenta.

As is well-known~\cite{Alkofer:2000wg,Bowman:2005zi}, the mass function is
strongly enhanced in the infrared.  This
is true even in the chiral limit: a clear demonstration of dynamical chiral
symmetry breaking.  The infrared value of around 350 MeV is consistent with
the constituent quark model.  $Z(q^2)$ is somewhat suppressed in the infrared.
A study on larger lattices reveals a flattening of both $M(q^2)$ and $Z(q^2)$
below around 500 MeV~\cite{Parappilly:2005ei}.  This is significant for
confinement, because an Euclidean propagator cannot have a point of inflexion
and adhere to reflection positivity~\cite{Roberts:2007ji}.

\begin{figure*}[t]
  \begin{center}
    \includegraphics[height=0.45\hsize,angle=90]{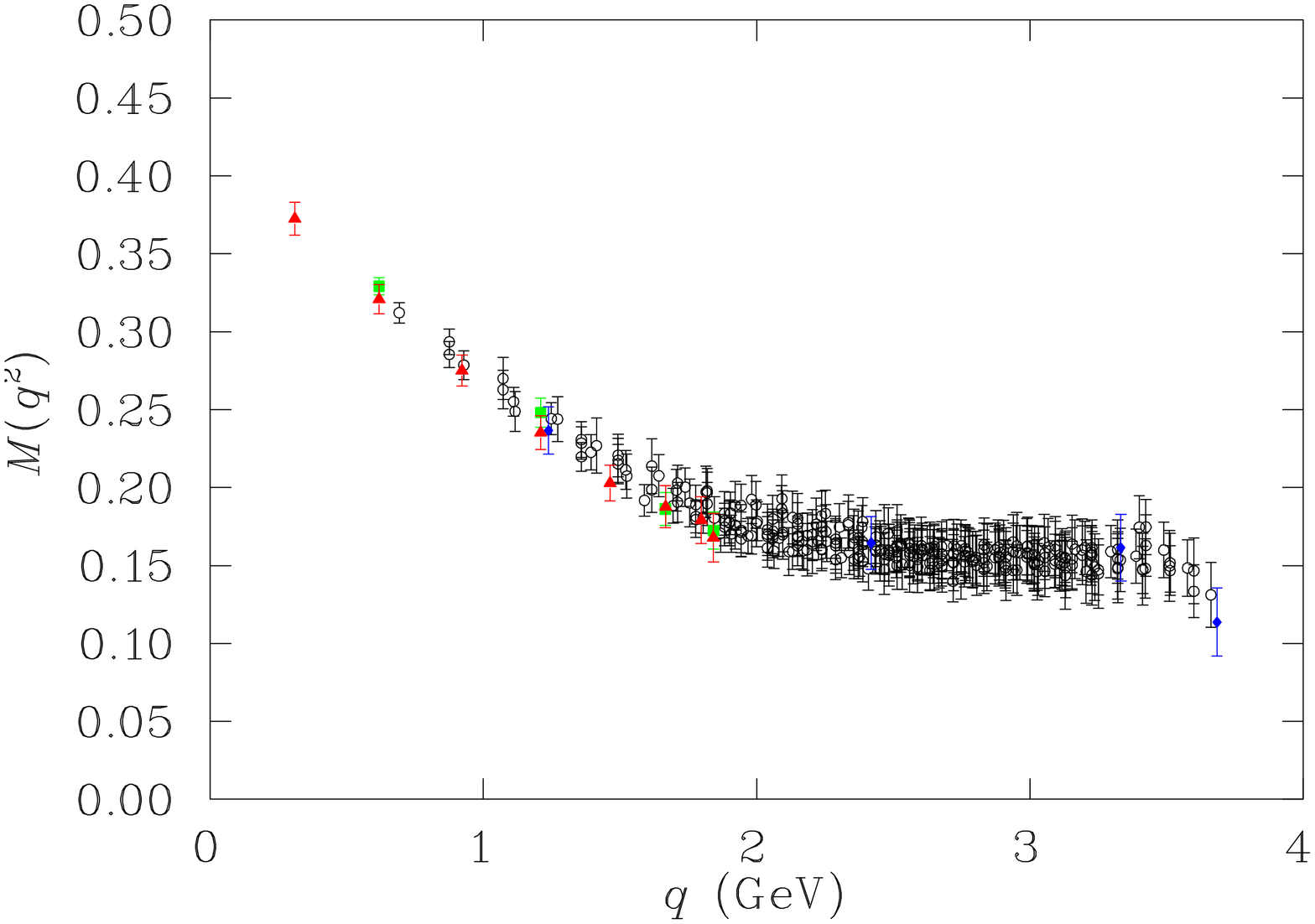}\hspace{5mm}
    \includegraphics[height=0.45\hsize,angle=90]{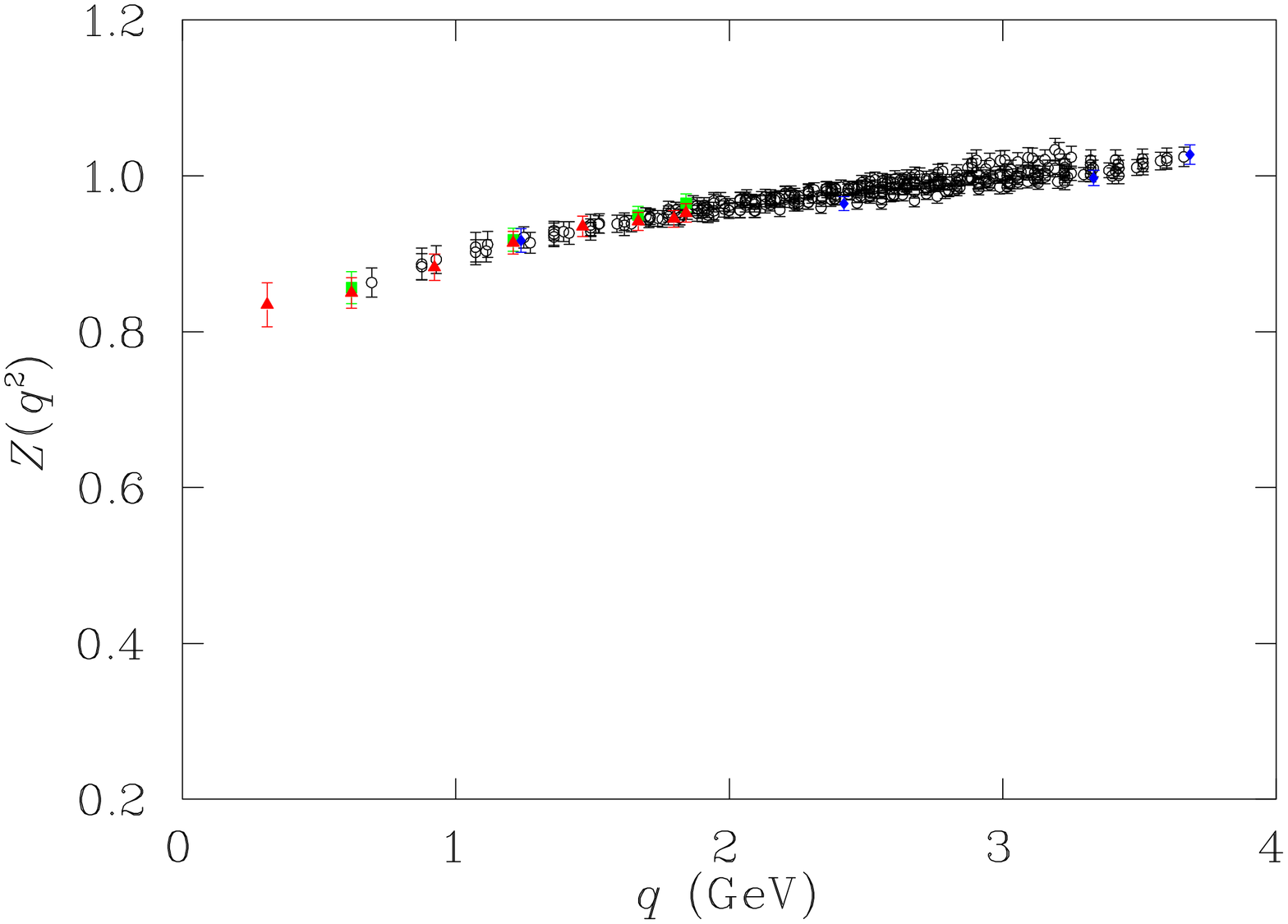}
    \caption{Landau gauge quark propagator for $m_0 a = 0.048$ following the
      removal of center vortices.  \dcsb\ still clearly dominates the mass
      function.  Both functions are somewhat flatter than on the full
      configurations.}
    \label{novort_M048}
  \end{center}
\end{figure*}

Figure~\ref{novort_M048} shows the mass and wave-function renormalization
functions after removing center vortices.  Mass generation associated with
dynamical chiral symmetry breaking is almost as strong after removing the
center vortices as it was before.  A roughening of the mass
function at large momenta suggests that the removal of center vortices
introduces significant noise into the gauge field configurations
giving rise to a larger effective mass.  $Z(p^2)$ is similarly weakly altered,
being slightly noisier and having less infrared suppression than on the
full configurations.

\begin{figure*}[t]
  \begin{center}
    \includegraphics[height=0.45\hsize,angle=90]{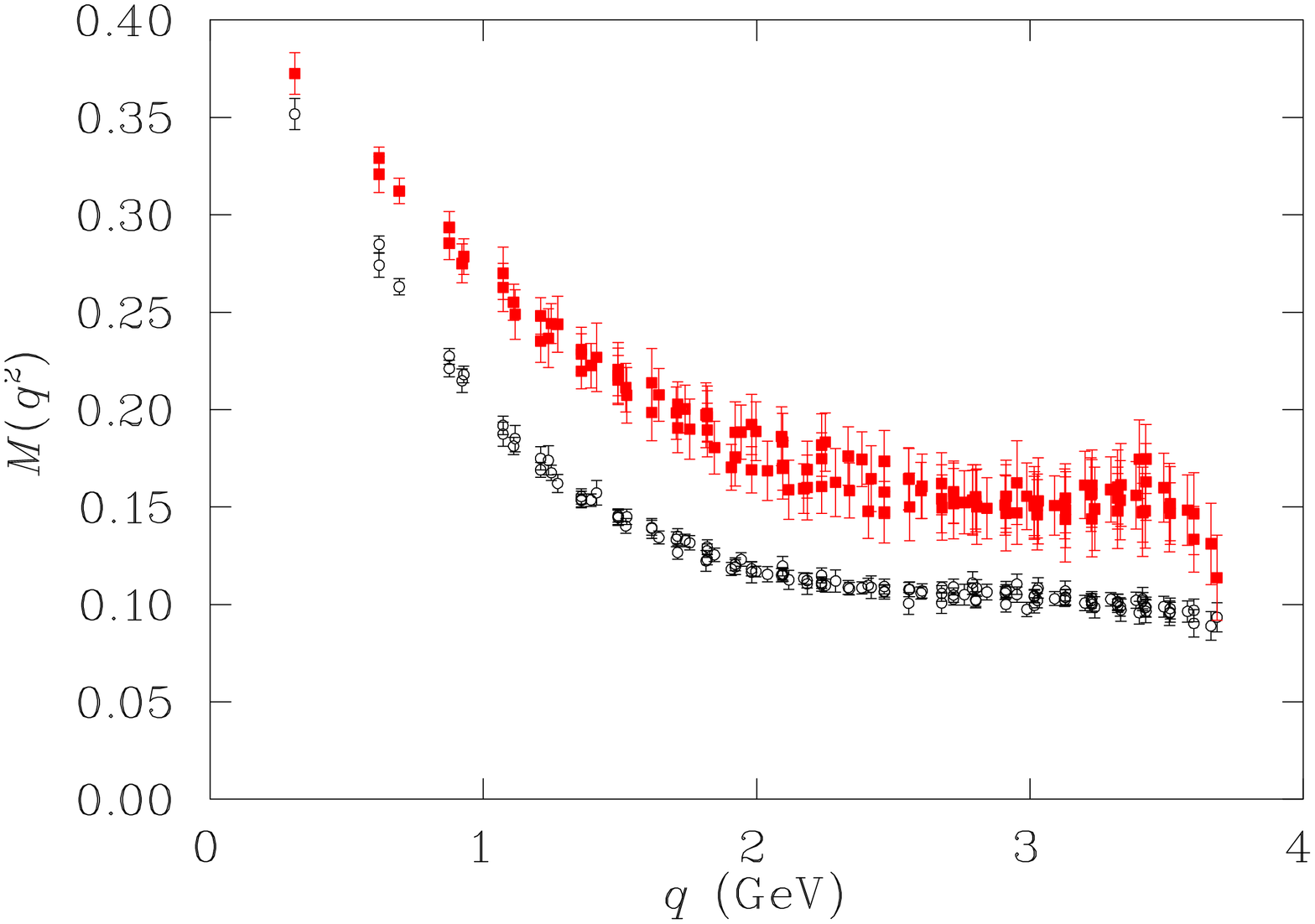}\hspace{5mm}
    \includegraphics[height=0.45\hsize,angle=90]{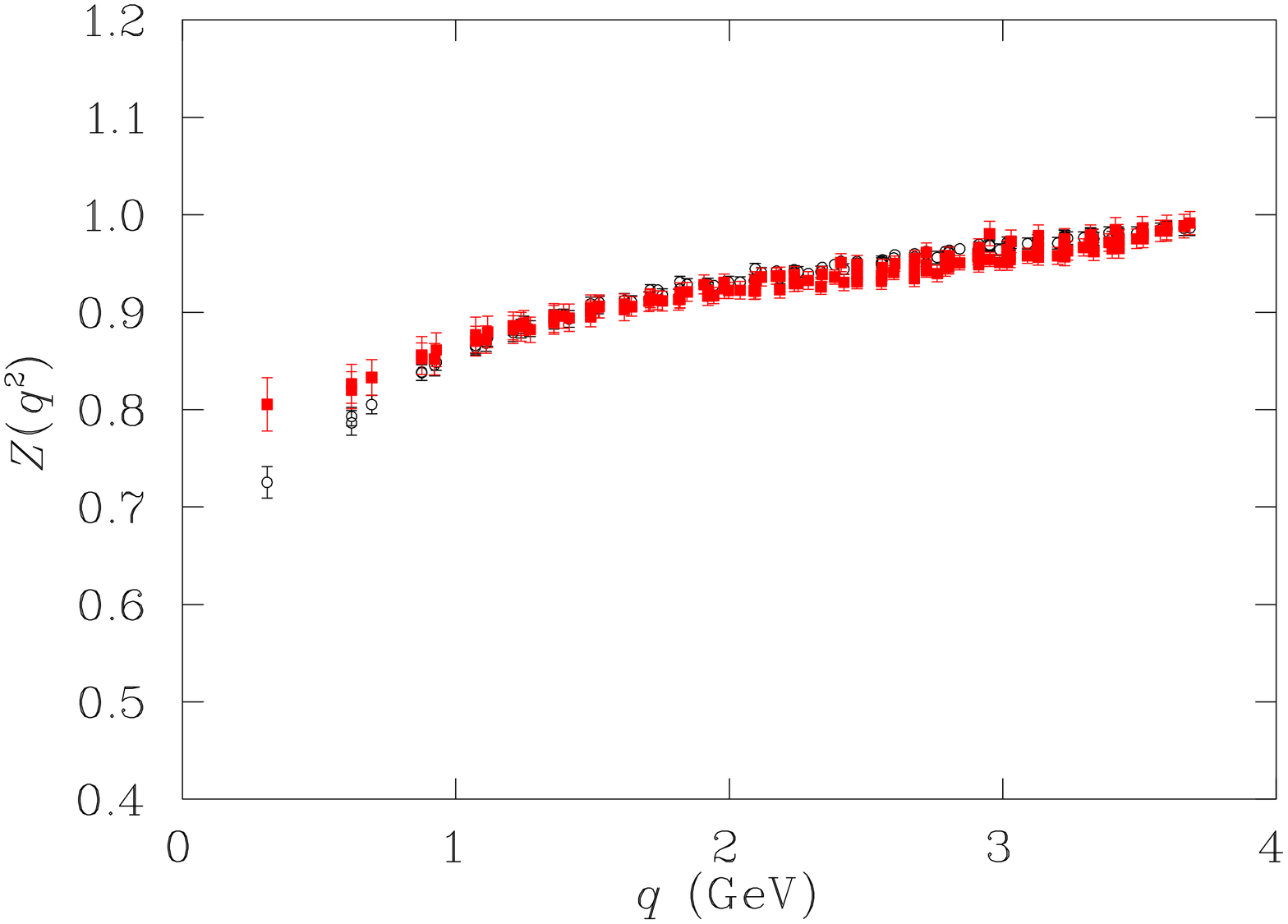}
    \caption{Landau gauge quark propagator for $m_0 a = 0.048$.  Open symbols
      denote the propagator obtained from the original gauge field
      configurations whereas the filled symbols denote the propagator
      following the removal of center vortices.  $Z(q^2)$ is renormalized to
      one at the largest accessible momentum point.}
    \label{vortcomp_m048}
  \end{center}
\end{figure*}

Figure~\ref{vortcomp_m048} shows a direct comparison of the quark propagator
on the full and vortex-removed configurations.  Data has been cylinder cut to
facilitate a detailed comparison.  The wave-function
renormalization function has been renormalized so that $Z(q^2) = 1$
at the largest momentum considered on the lattice.  Only below about 1 GeV is
there any significant difference between the full and vortex-removed results.
It is possible that the removal of
center vortices has caused $Z(p^2)$ to straighten out, which could restore
reflection positivity and hence be a sign of deconfinement.  

\begin{figure}[t]
  \begin{center}
    \includegraphics[height=0.9\hsize,angle=90]{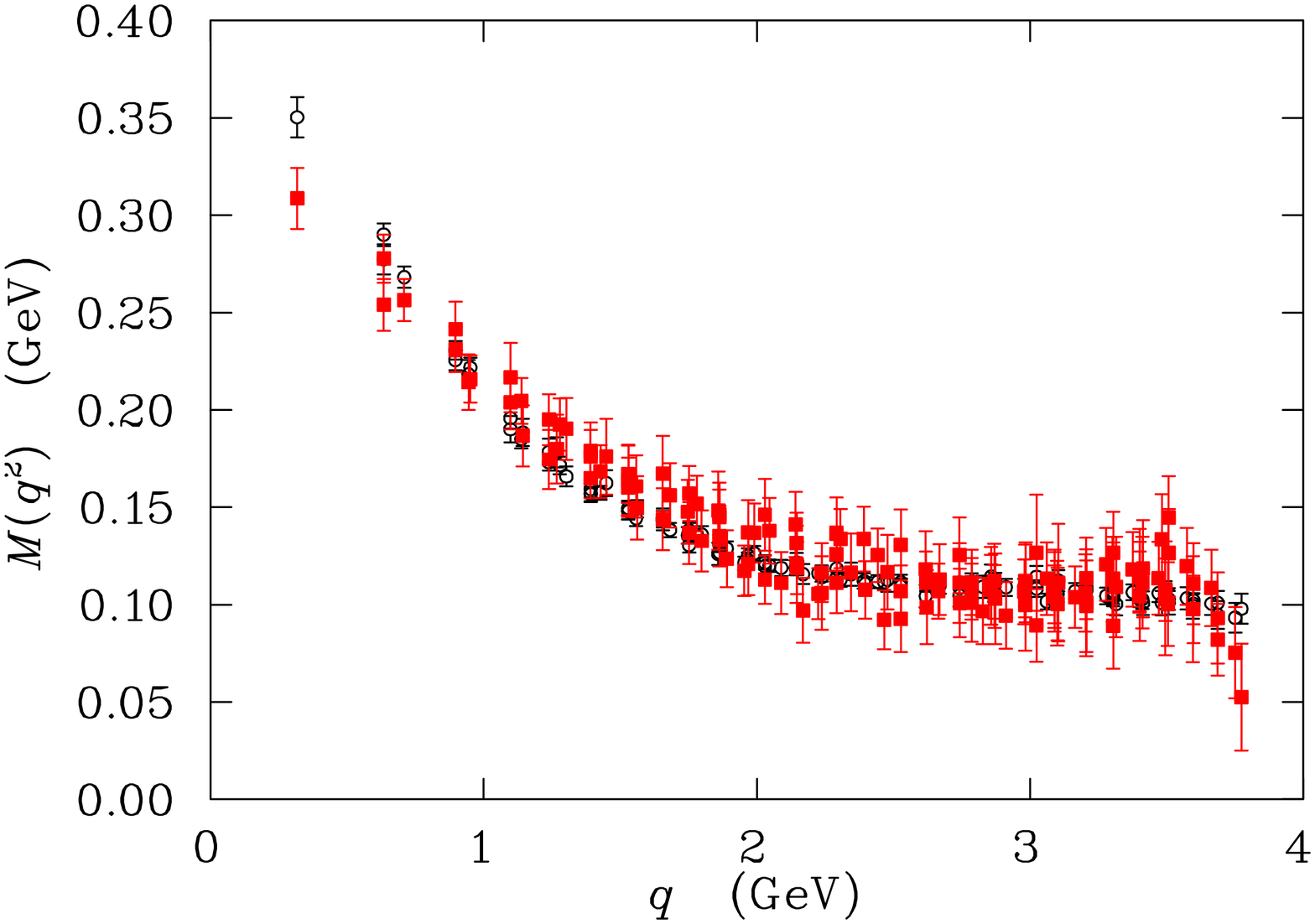}
    \caption{The Landau gauge quark propagator with $m_0 a = 0.048$ from
      the original configurations (open symbols) is compared with the
      propagator obtained from the vortex-removed configurations with
      $m_0 a = 0.024$ (filled symbols) selected to match the
      renormalized quark mass in the ultraviolet regime.}
    \label{MassFm048r024}
  \end{center}
\end{figure}

The mass function does not undergo a multiplicative renormalization,
as described in Sec.~\ref{sec:quarkprop}.  However, removing the
center vortices has significantly increased the running mass as
displayed in the ultraviolet regime of the mass function.  An
alternative analysis is to compare full and vortex-removed results
with bare quark masses adjusted to provide matched running quark
masses. Figure~\ref{MassFm048r024} illustrates the persistent nature
of the mass function under vortex removal.  In this case we see that
removing the vortices suppresses $M(q^2)$ near zero four-momentum by
about $15\%$ compared to the full configurations, weakening -- but by
no means eliminating -- strong infrared enhancement.  Either way,
there is still plenty of dynamical mass generation.

\begin{figure*}[t]
  \begin{center}
    \includegraphics[height=0.45\hsize,angle=90]{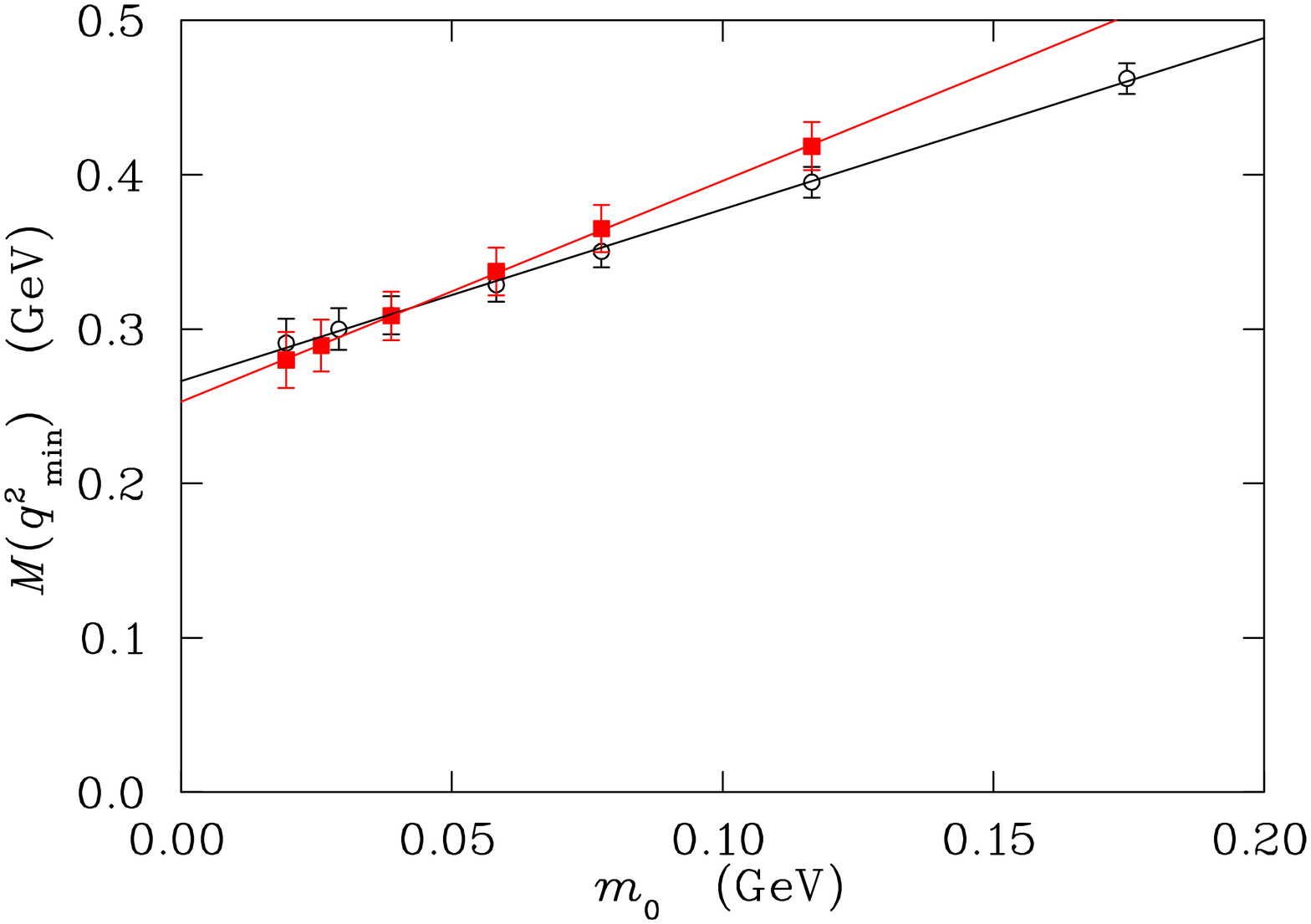}\hspace{5mm}
    \includegraphics[height=0.45\hsize,angle=90]{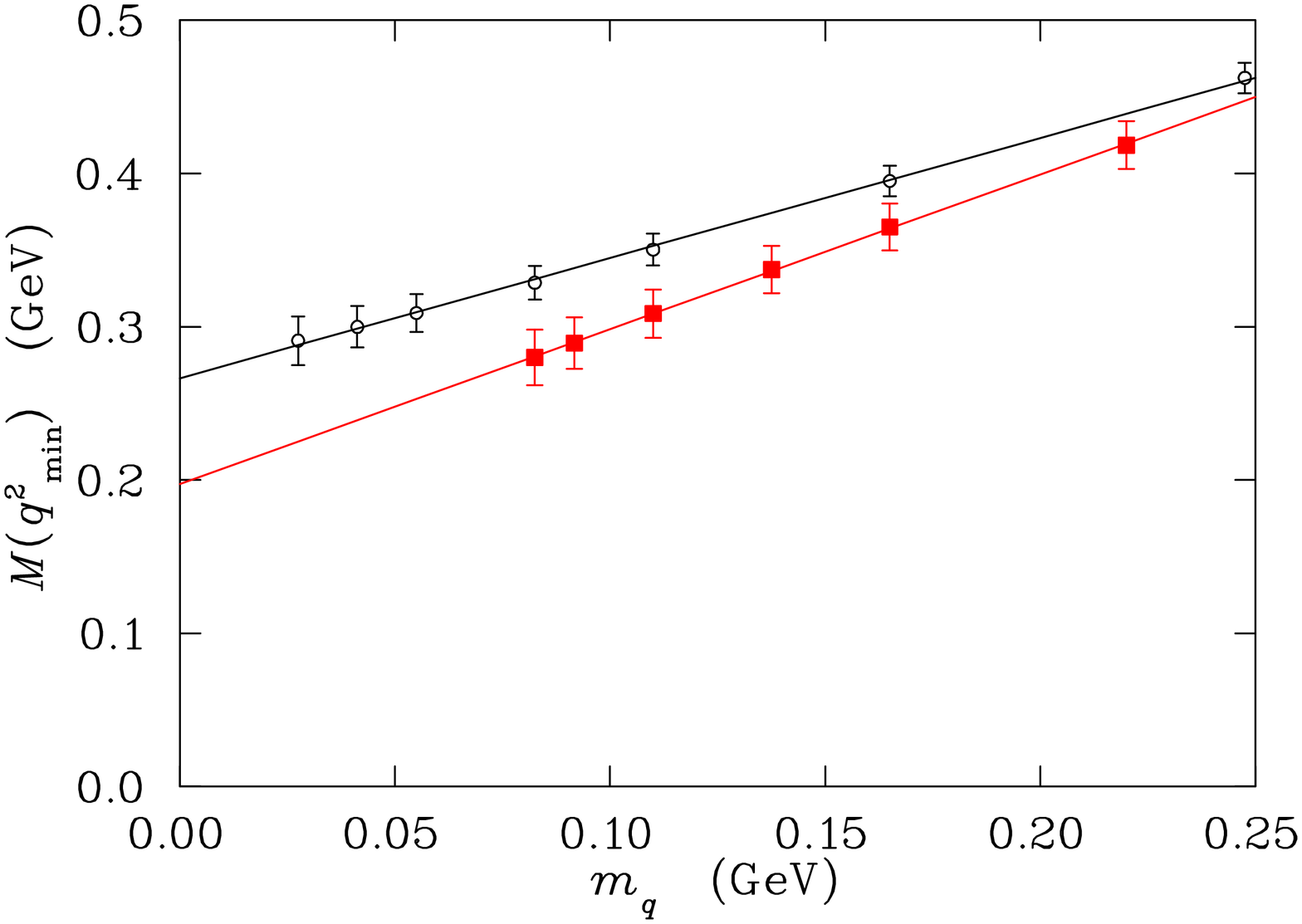}
    \caption{Mass function at the smallest nontrivial momentum
      available on our lattice, $M(q^2_{\rm min})$, for a variety of
      bare quark masses, $m_0$.  Open symbols denote the mass function
      obtained from the original gauge field configurations whereas
      the filled symbols denote the mass function following the
      removal of center vortices.  Original and vortex-removed results
      are compared for equal bare quark masses (left) and equal
      renormalized quark masses, $m_q$, at $q = 3.0$ GeV (right).}
    \label{extrap}
  \end{center}
\end{figure*}

To further explore the infrared nature of the quark mass function we
turn our focus to the value of the mass function at the smallest nontrivial
momentum available on our lattice, $q^2_{\rm min} = 0.10$ GeV${}^2$.
Figure~\ref{extrap} compares the mass function at $q^2_{\rm min}$ for
a variety of bare quark masses, $m_0$.  In the left-hand plot,
$M(q^2_{\rm min})$ is compared directly without an adjustment of the
bare quark mass, whereas the right hand plot compares $M(q^2_{\rm
  min})$ with the input bare masses adjusted to provide similar
renormalized quark masses, $m_q$ at $q = 3.0$ GeV.  Linear fits are
sufficient to describe the data and indicate significant dynamical
mass generation in the chiral limit.

In the early days of Dyson-Schwinger studies of QCD, \dcsb\ was
attributed to the interaction strength in the quark sector provided by
an effective 1-gluon exchange.  Although it is now clear
\cite{Alkofer:2004it} that vertex corrections play an important role
in the quark IR sector, it is unlikely that a loss of gluonic
interaction strength (as displayed by the full gluon propagator) goes
unnoticed with regard to \dcsb.  On the basis of our findings above,
it is therefore interesting to study the effect of center vortex
removal on the Landau gauge gluon propagator.

\begin{figure}[t]
  \begin{center}
    \includegraphics[height=0.8\hsize,angle=90]{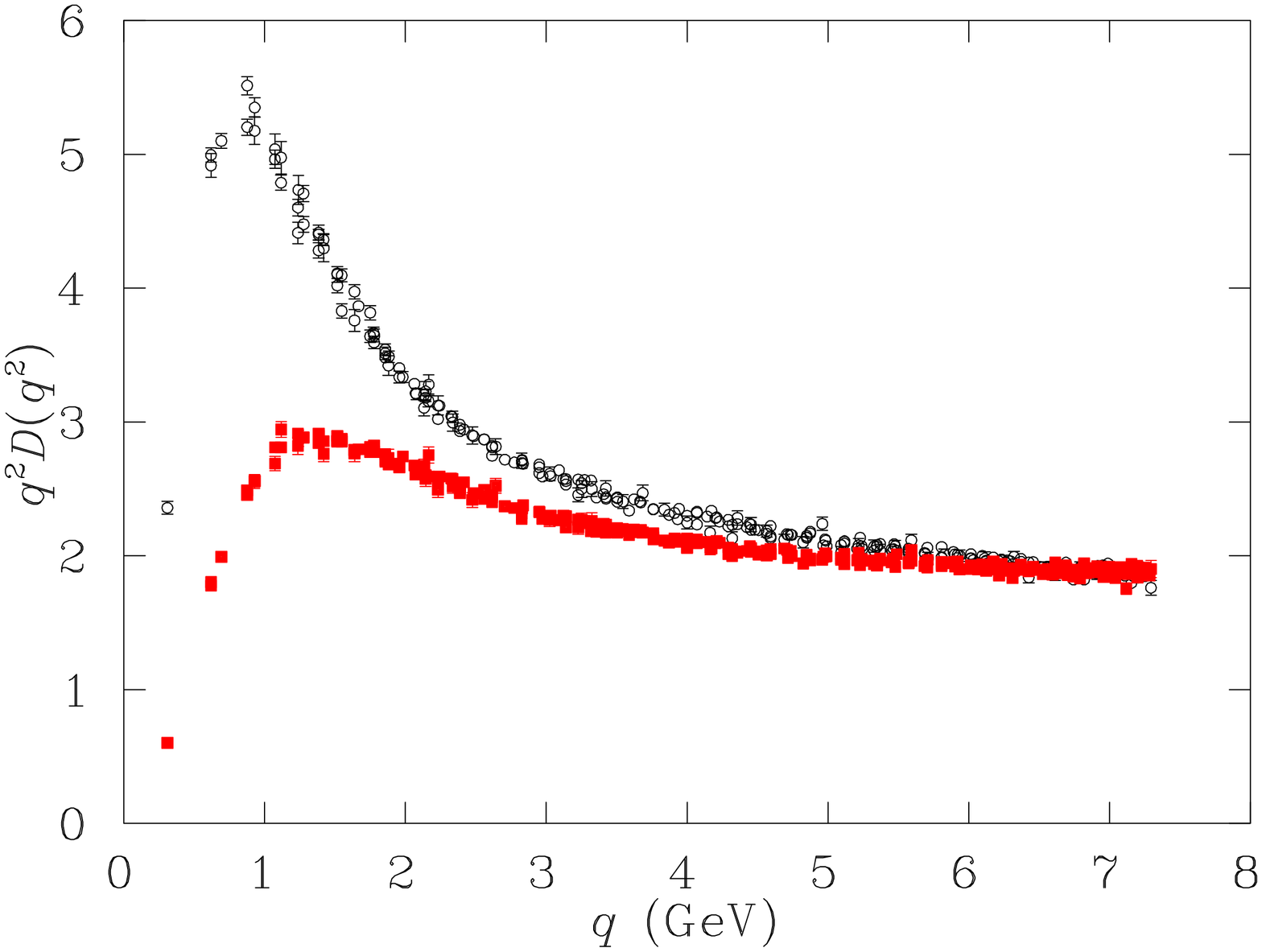}
    \caption{The gluon propagator multiplied by $q^2$ (gluon dressing function)
      such that the large momentum value approaches a constant.  The data
      presented here have been cylinder cut.  Open symbols
      denote results from the original untouched gauge fields while full
      symbols report the propagator after removing center vortices. }
    \label{vortcomp_gluon}
  \end{center}
\end{figure}

As far as the full gluon propagator, $D(q^2)$, is concerned, it is
known to be infrared enhanced, but finite at zero
four-momentum~\cite{Bonnet:2001uh,Sternbeck:2007ug,Cucchieri:2007rg,Bogolubsky:2009dc}. 
This can be seen in Fig.~\ref{vortcomp_gluon} from the gluon dressing
function, $q^2 D(q^2)$ of the Landau-gauge gluon propagator.  At high
momenta, the dressing function logarithmically decreases with
momentum, while it is enhanced at intermediate momenta with a maximum
near 1 GeV.  The turn over indicates a violation of positivity, as
explicitly shown in
Ref.~\cite{Cucchieri:2004mf,Sternbeck:2006cg,Bowman:2007du}.  The same
picture was found in full QCD with light sea
quarks~\cite{Bowman:2004jm,Bowman:2007du}.  Also shown in
Fig.~\ref{vortcomp_gluon} is the dressing function upon vortex
removal.  As for the gauge group
SU(2)~\cite{Langfeld:2001cz,Gattnar:2004bf}, we find that the infrared
enhancement is largely suppressed when center vortices are
removed. This is particularly remarkable in the light of our previous
findings: vortex removal strongly reduces the gluonic interactions
strength, but dynamical mass generation is largely unaffected.

\begin{figure}[t]
  \begin{center}
    \includegraphics[height=0.9\hsize,angle=90]{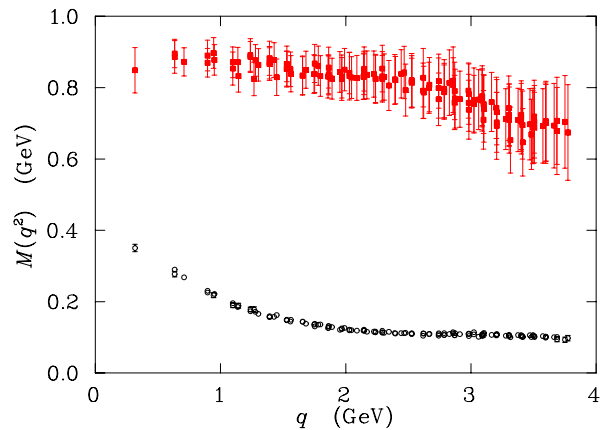}
    \caption{Landau gauge quark propagator for $m_0 a = 0.048$.  Open symbols
      denote the propagator obtained from the original gauge field
      configurations whereas the filled symbols denote the propagator
      following the removal of center vortices identified in Laplacian
      Centre Gauge.}
    \label{LCGmassFun}
  \end{center}
\end{figure}

On a final note, we have also investigated the role of center vortices
defined using Laplacian Center Gauge (LCG)~\cite{deForcrand:2000pg},
where the Laplacian gauge construction removes the gauge-fixing
ambiguity.  While vortices defined this way do account for the full
string tension, the vortex density diverges in the continuum
limit~\cite{Langfeld:2003ev}.  On a practical side, there is an
abundance of LCG vortices in the vacuum.  Upon removing these vortices
the configurations become extremely rough.  The mass function revealed
following LCG-vortex removal bares little resemblance to the original
mass function.  It is dominated by noise at all distance scales as
illustrated in Fig.~\ref{LCGmassFun}, where the scale has been
adjusted to accommodate the results.

\section{Discussion}

Using the SU(3) vortex picture defined by means of the mesonic version
of the Maximal Center Gauge~\cite{Faber:1999sq,Montero:1999by} allows
us to switch to non-confining QCD upon vortex removal. As for the
gauge group SU(2), an inspection of the gluonic dressing function
shows a strong decline of gluonic interaction strength. This alone
stirs the expectation that chiral symmetry might be restored as well.

By contrast to this expectation, we find that the removal of center
vortices from our configurations has done little to interfere with
chiral symmetry breaking, as seen by the persistent infrared
enhancement of the quark mass function.  The analogy to the SU(2)
gauge group \cite{Bowman:2008qd} ends here: in SU(2), vortex removal
implies the restoration of chiral
symmetry~\cite{deForcrand:1999ms,Langfeld:2003ev,Gattnar:2004gx}.

We do stress that the key to SU(3) center vortex matter might not have
been found yet.  This is most obvious from the impact of the presently
defined SU(3) vortices on the static quark potential: the string
tension vanishes on vortex-removed configurations, but only of the
order two thirds of the full string tension is recovered on
vortex-only configurations. 

Whether the phenomenon of \dcsb\ disentangles from quark confinement
for the SU(3) gauge group (in contrast to the SU(2) case) or whether
an improved definition of SU(3) vortex texture is yet to be discovered
needs further investigation. Yet our findings offer the intriguing
possibility to separately trace out the impact of confinement and
the impact of \dcsb\ on hadronic observables, at least in SU(3).


\acknowledgments 

POB acknowledges fruitful discussions with Craig Roberts and Peter
Tandy.  This research was undertaken on the NCI National Facility in
Canberra, Australia, which is supported by the Australian Commonwealth
Government.  We also acknowledge eResearch SA for generous grants of
supercomputing time which have enabled this project.  This work was
supported by the Australian Research Council.  The work of POB was
supported by the Marsden Fund of the RSNZ.

\bibliography{references}

\end{document}